\documentclass[referee]{aa} 
\usepackage[comma,authoryear]{natbib}
\usepackage{graphics}
\usepackage[latin1]{inputenc} 

\begin{document}

 \title{Reconstructing the solar integrated radial velocity using MDI/SOHO}

   \titlerunning{}

   \author{N. Meunier \inst{1}, A.-M. Lagrange \inst{1}, M. Desort \inst{1} 
	  }
   \authorrunning{Meunier et al.}

   \institute{
Laboratoire d'Astrophysique de Grenoble, Observatoire de Grenoble, Universit\'e Joseph Fourier, CNRS, UMR 5571, 38041 Grenoble Cedex 09, France\\
  \email{nadege.meunier@obs.ujf-grenoble.fr}
             }

\offprints{N. Meunier}

   \date{Received 5 February 2010 ; Accepted 12 May 2010}

\abstract{Searches for exoplanets with radial velocity techniques are increasingly sensitive to stellar activity. It is therefore crucial to characterize how this activity influences radial velocity measurements in their study of the detectability of planets in these conditions. }
{In a previous work we simulated the impact of spots and plages on the radial velocity of the Sun. Our objective is to compare this simulation with the observed radial velocity of the Sun for the same period. }
{We use Dopplergrams and magnetograms obtained by MDI/SOHO over one solar cycle to reconstruct the solar integrated radial velocity in the Ni line 6768 \AA. We also characterize the relation between the velocity and the local magnetic field to interpret our results. }
{We obtain a stronger redshift in places where the local magnetic field is larger (and as a consequence for larger magnetic structures): hence we find a higher attenuation of the convective blueshift in plages than in the network. Our results are compatible with an attenuation of this blueshift by about 50\% when averaged over plages and network. We obtain an integrated radial velocity with an amplitude over the solar cycle of about 8 m/s, with small-scale variations similar to the results of the simulation, once they are scaled to the Ni line. }
{The observed solar integrated radial velocity agrees with the result of the simulation made in  our previous work within 30\%, which validates this simulation. The observed amplitude confirms that the impact of the convective blueshift attenuation in magnetic regions will be critical to detect Earth-mass planets in the habitable zone around solar-like stars. }


\maketitle

\section{Introduction}

Searches for exoplanets with radial velocity (hereafter RV) techniques will be increasingly sensitive to stellar activity as the sensitivity of instruments is improved \cite[such as Espresso on the VLT or Codex on the E-ELT][]{dodorico07}. It is therefore crucial to understand how this activity influences the RV measurements. 
In our previous works, we simulated the influence of spots \cite[][hereafter Paper I]{lagrange09} and plages \cite[][hereafter Paper II]{meunier09} on the radial velocity (hereafter RV) for the Sun seen equator-on. The simulation covered a solar cycle with an excellent temporal cadence. Our study considered the photometric contribution of spots and plages and also the attenuation of the convective blueshift in magnetic regions (Paper II). This last effect appeared to be dominant, with a long-term amplitude over the solar cycle of about 8~m/s, while the photometric contributions produced a rms of 0.08 m/s during cycle minimum and of 0.42 m/s during cycle maximum (0.33~m/s over the whole cycle), with peak-to-peak amplitudes of a few m/s. We studied how these RV variations influence the detectability of an Earth-mass planet in the habitable zone and found that it made this detection very difficult. 

Most existing observations of the solar RV do not cover our data set. Neither are the investigated lines well characterized in terms of convective blueshift, which makes them difficult to compare with simulations. \cite{mcmillan93} measured the solar RV with the moon light observed in the violet part of the spectrum for the period 1987--1992. They found solar RV variations that are significantly smaller than our simulation. For the period 1983--1992, \cite{DP94} measured the solar RV in the 2.3~$\mu$m domain while \cite{jimenez86} measured it in the potassium line at 769.9~nm \cite[][]{brookes78} during cycle 21 (1976--1984). Both have found an amplitude over the cycle of about 30~m/s. The potassium line is a strong line however, for which a particularly high convective blueshift is not expected.  
The GOLF \cite[][]{gabriel95} instrument on SOHO provides integrated measurements of the solar RV in the Na D1 and D2 lines. 
Unfortunately, the planned calibration procedure could not be applied to the GOLF observations because of an instrumental failure early in the mission. Therefore the long-term variations of the RV signal and its amplitudes are unreliable (P. Boumier, private communication). We note that after a few years of operation, a comparison has been made between the MDI and GOLF signal \cite[][]{henney98} but focused on the high temporal frequency component and not on the long-term signal. Some attempts to calibrate the long term variations a posteriori, mostly due to the spacecraft orbit, have been made \cite[][]{ulrich00} but the resulting signal still shows some strong instrumental residuals that are not compatible with the amplitude we wish to test. 


This is why precise RV observations of the Sun in integrated light are necessary for testing our simulations, in particular for the convective blueshift contribution. It would be best to directly observe the Sun in integrated light over an extended domain of the spectrum with an instrument and data analysis similar to those used to measure stellar RV variations. This is a long-term project though. The aim of this paper is to characterize the solar RV variations with spatially resolved observations by MDI/SOHO in a single spectral line (Ni line at 6768 \AA). The use of an imaging instrument also allows us to identify the contribution of solar activity to the solar integrated velocities, in contrast with an instrument such as GOLF. We reconstruct the integrated RV variations (neglecting the photometric component of spots and plages simulated in Paper I and II) and characterize the behavior of the velocity variations measured in the Ni line with respect to the magnetic field. Our observations are compared to the simulations performed in Paper II scaled to the expected behavior of that line.  

The input data and data analysis are described in Sect.~2. We describe how we determine the zero velocity for each data point in the time series and estimate the uncertainty of our measurements. The local velocity - magnetic field relation derived from the analysis of the Ni line is presented in Sect. 3. This relation allows the interpretation of the integrated RV variations. We also discuss the origin of the velocity associated to magnetic regions in term of convective blueshift attenuation and downflows. Then we present the integrated RV in Sect.~4 and compare it to the results of the simulation of the convective blueshift attenuation made in Paper II scaled to the Ni line expected behavior. The comparison is made at different time-scales. We conclude in Sect.~5.

\section{Observation and data analysis}

\subsection{Input data}

We use a series of MDI/SOHO Dopplergrams and magnetograms \cite[][]{Smdi95} covering a solar cycle between 1996 May 5 (Julian day 2450209) and 2007 October 7 (Julian day 2454380), for the Ni 6768$\:$ line. The intensity in the line core is about 0.36 times the continuum intensity. Dopplergrams are maps of the line-of-sight velocity (positive velocities correspond to a redshift), and are dominated by the large-scale motions (mostly the rotation) and small-scale motions mostly due to granulation and supergranulation. An example is shown in Fig.~\ref{map} (uppper left panel). Magnetograms are maps of the line-of-sight magnetic field averaged over each pixel, i.e. flux maps. 
We selected Dopplergrams obtained on the same days as the magnetograms analyzed in Paper II and separated in time by less than six hours, which leads to 3181 data points (out of which 74\% corresponds to a separation less than one hour). 

The MDI data observations are made only for the Ni 6768$\:$ line.
To derive a convective blueshift for our entire spectral range, we used the empirical relation between the convective blueshift and the line depths \cite[][]{gray09} in Paper II. The average convective blueshift was also weighted by the relative contributions of various lines to the correlation function used to compute radial velocities, to produce a convective blueshift valid for the entire wavelength range. Given the depth of the Ni line, and an attenuation by a factor two-thirds \cite[][]{bs90} of the convective blueshift in magnetic regions as in Paper II, we expect a vertical velocity $\Delta$V in the magnetic regions for that line of 227~m/s (instead of 190~m/s for the entire spectrum). This value is used in Sect.~4 to scale the results of the simulation of Paper II to compare it with our specific conditions. 
 
\subsection{Dopplergram analysis}

\begin{figure*} 
\includegraphics{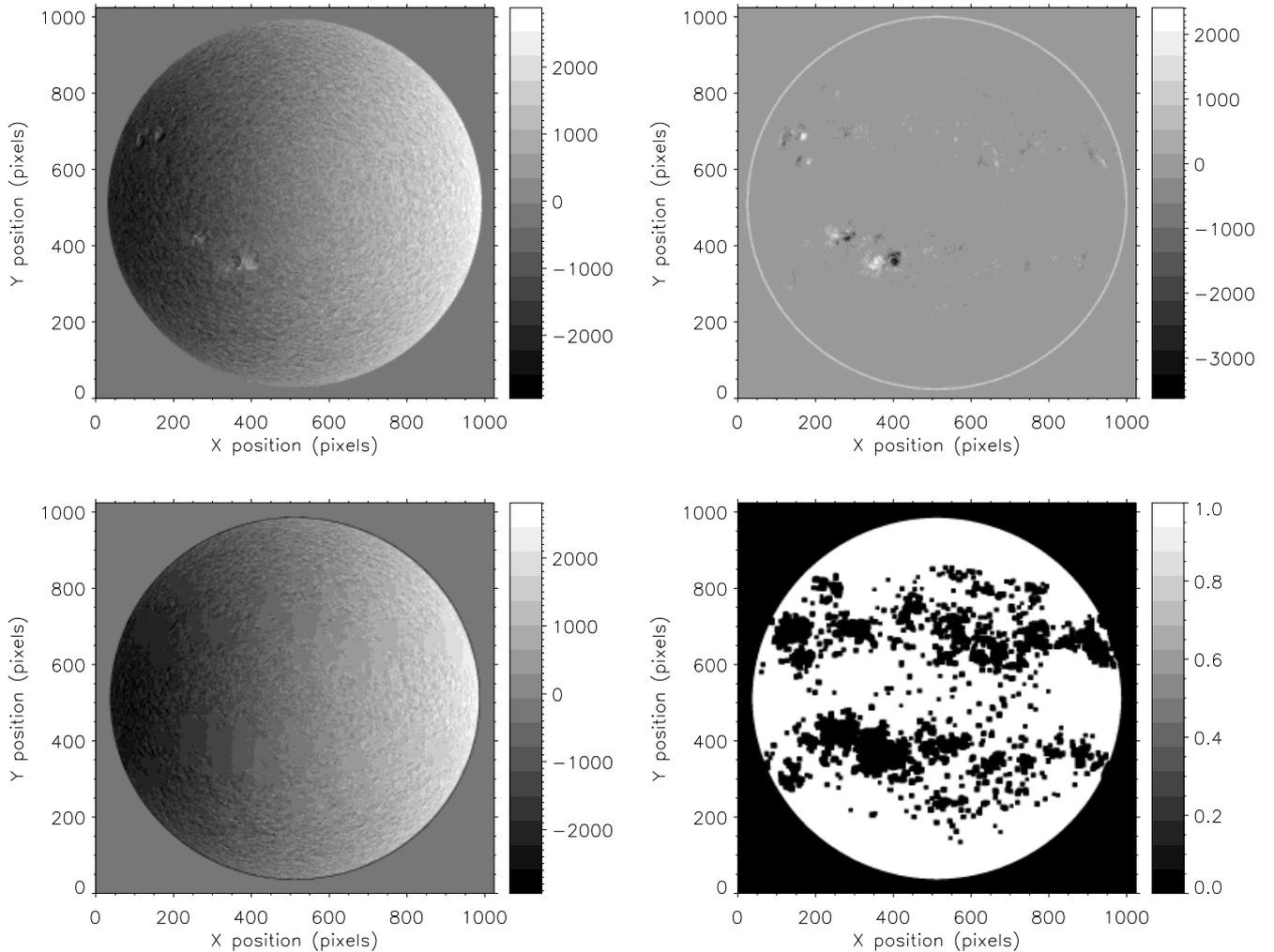}
\caption{{\it Upper left panel}: Dopplergram for 2000 May 16 (in m/s). {\it Upper right panel}: Magnetogram for the same day (in Gauss). {\it Lower left panel}: Dopplergram after interpolation (in m/s). {\it Lower right panel}: Mask derived from the magnetogram (on the disk, 0 where the Dopplergram is interpolated). }
\label{map}
\end{figure*}

\begin{figure} 
\includegraphics{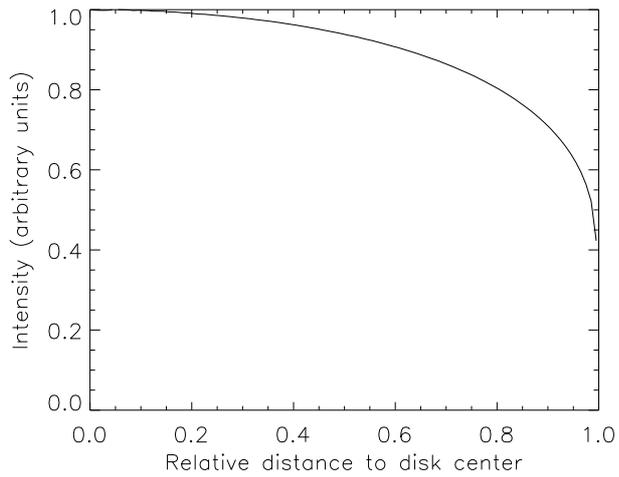}
\caption{Limb-darkening deduced from 78 MDI intensity maps during a very quiet period.}
\label{assomb}
\end{figure}

\begin{figure} 
\includegraphics{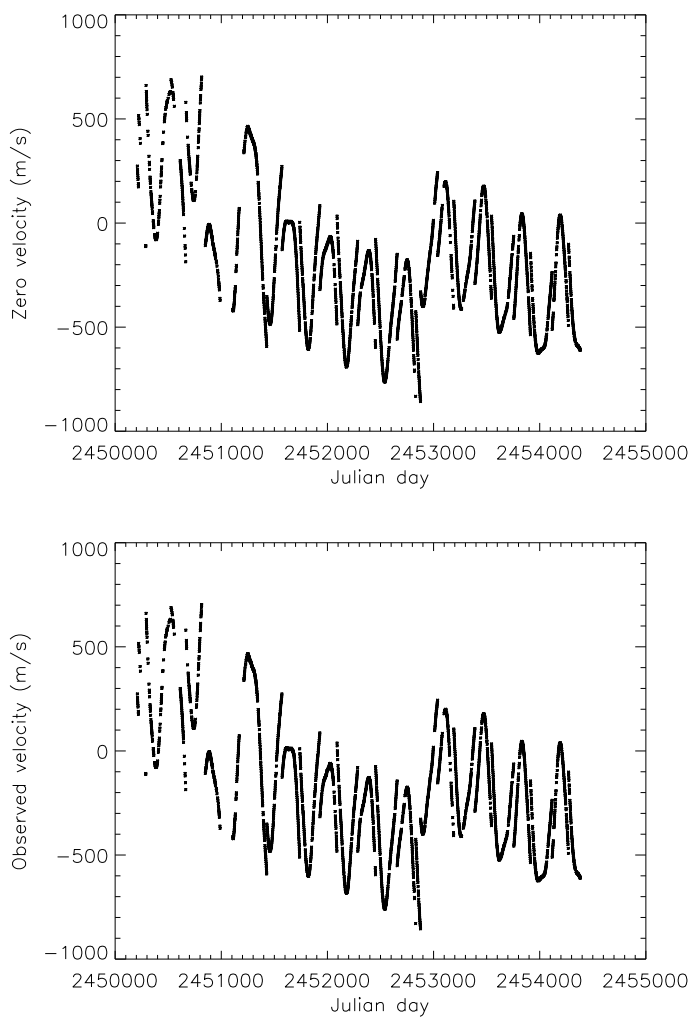}
\caption{{\it Upper panel}: Zero velocity ($V_{\rm zero}$) determined on interpolated Dopplergrams. {\it Lower panel}: Velocity ($V_{\rm obs}$) integrated over Dopplergrams. }
\label{vall}
\end{figure}

Radial velocities measured by MDI do not present a good long-term stability \cite[e.g.][]{beck98,evans98}, mostly due to the orbit of the spacecraft (amplitude of about 500 m/s) and semi-annual changes in the wave plate angles, producing a typical drift of 1~m/s per day and steps of several hundreds of m/s typically twice a year. 
The zero of MDI Dopplergrams is not sufficiently good to directly derive an integrated radial velocity from the observations and we therefore need to define this zero using the data themselves. 

The principle of our approach is the following (the details will be given below). We assume that the quiet Sun provides an appropriate zero velocity map. Using the magnetogram observed at a given time, we define a mask by selecting the magnetic regions. This mask is applied to the corresponding (in time) Dopplergram to identify regions where the velocity is affected by magnetic activity. The velocity in those regions will have to be interpolated from the unmasked areas (which represent the quiet Sun). Then we average the velocities over the disk, which produces a reference for the radial velocity variations. 

The masks were derived from each magnetogram. First, we corrected the magnetogram for the rotation: if the Dopplergram and magnetogram are not obtained strictly simultaneously, the rotation induces a difference in position between a given region on the two images. A threshold of 50~G was then applied on the slightly smoothed magnetograms to define magnetic regions, producing a binary mask. This mask was then enlarged (10 pixels around each structure). This allowed us to eliminate most flows caused by magnetic regions without eliminating too many pixels. An example of such a mask is shown in Fig.~\ref{map} (lower right panel).    

The mask was then used to define zones in the Dopplergram where the velocity is interpolated from quiet Sun values around the mask. Because we aim to determine a zero velocity that would correspond to the quiet Sun, we built a Dopplergram that would be similar to a quiet Sun Dopplergram. For each pixel we used a linear interpolation between four pixels after smoothing smoothing the Dopplergrams. These four pixels were the closest pixels outside the mask that are respectively on the right and left (same line) and above and below (along the same column) the considered pixel. Figure~\ref{map} (lower left panel) shows a Dopplergram interpolated in this way. 

The integrated velocity is derived from the actual Dopplergram (thus providing $V_{\rm obs}$) and from the interpolated reference Dopplergram (thus providing $V_{\rm zero}$) after a weighting by the center-to-limb darkening shown in Fig.~\ref{assomb}. If we were actually observing the Sun in integrated light directly, the velocity corresponding to each pixel on the Sun would indeed be weighted by the corresponding intensity. 
Figure~\ref{vall} shows the complete time series for $V_{\rm zero}$ and $V_{\rm obs}$. 
Note that in principle it should be possible to use the  spacecraft velocity as reference. However, the use of the spacecraft radial velocity shows that some strong residuals, which are related to the instrument, are still present. 

\subsection{Uncertainties}

\begin{figure} 
\includegraphics{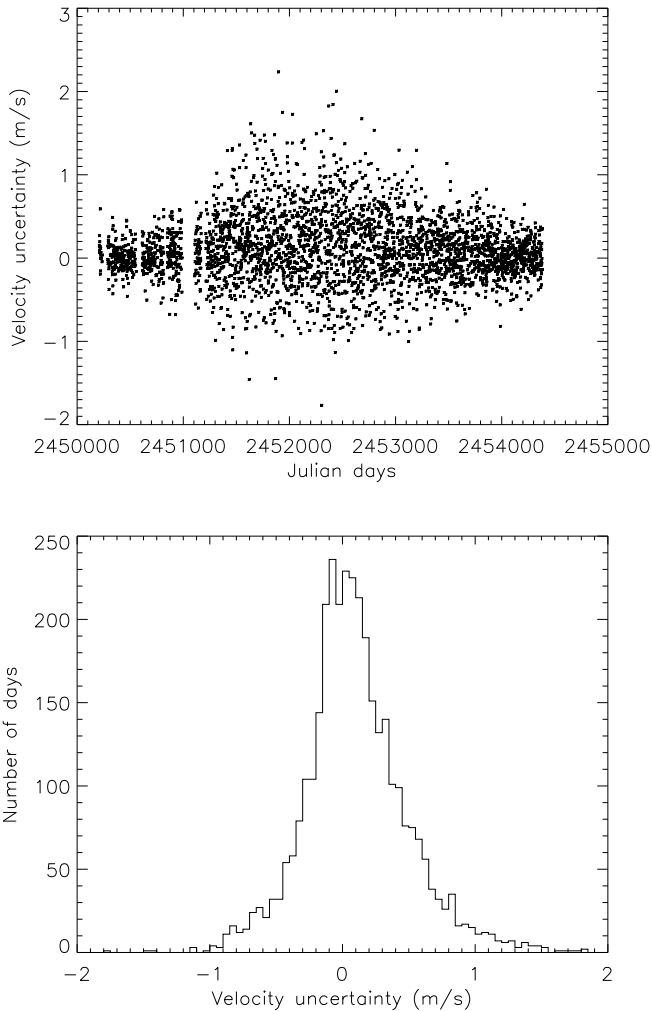}
\caption{{\it Upper panel}: Estimated errors of the velocity versus time. {\it Lower panel}: Distribution of the estimated errors. }
\label{err}
\end{figure}

The identified sources of uncertainties are the statistical noise in the data caused by the actual motions at the granular and supergranular scales, the Evershed effect around sunspots, the interpolation made to estimate the zero velocity in the regions of magnetic activity and the exact definition of the mask. We study the relative importance of these factors.  

To test the zero of the velocity, we considered a very quiet Dopplergram, for which we assumed the observed velocity to be our reference $V_{\rm ref}$ (i.e. we assume that the magnetic regions do not significantly influence the average velocity in that case). Then we applied to that same Dopplergram the mask derived for each data point in our time series, interpolated, and computed the radial velocity $V(t)$ of the corrected Dopplergrams. Ideally the two should be identical. Figure~\ref{err} shows $V(t)-V_{\rm ref}$ for the entire time series (3415 data points) and the distribution of the values. The rms is 0.4~m/s, and is lower during cycle minimum (0.20--0.25 m/s) than during cycle maximum (0.5 m/s). Part of this dispersion is due to the interpolation. 
The dispersion in velocity over the solar disk is also a source of uncertainty (and is included in the previous measurement). After removing large-scale motions (solar rotation and meridional circulation), the rms is 220 m/s,  which leads to an average uncertainty of about 0.3 m/s considering the number of pixels. This is not much below 0.4 m/s, which shows that the contribution of the interpolation to the uncertainty is small (probably not more than 0.2--0.3 m/s during high-activity periods). 
Averaging the Dopplergrams in time (for example one hour) does not help in reducing this noise because supergranulation dominates on longer time scales so that the rms remains large. 

We also built similar time series using other quiet Dopplergrams : they lead to a similar dispersion. However, the exact values are different (two of these series are not correlated), so this time series cannot be used to correct the RV signal for instance, only the dispersion is of interest.  
Note that part of the uncertainty may arise because even during the quiet periods that we consider, there are still a few magnetic structures: depending on the mask some may be eliminated or not, which adds an uncertainty to this error estimate itself.   

We also tested the impact of the mask. As described in the previous section, the main parameters to derive the mask are the threshold used to define magnetic regions, and the dilatation of the resulting mask by 10 pixels around each structure. This last effect dominates the first one because the magnetic regions are defined with a much better precision. It is difficult to use a lower value, because the Evershed effect would then significantly affect the zero, especially for large active regions. We therefore tested slightly different values above 10 pixels, keeping in mind that when considering too high values the resulting pixels will become too few to perform a reliable interpolation. Then we compute the rms of the difference between pairs of series. It appears to be difficult to derive an exact value of the uncertainties, as the rms varies depending on the pair, but values lie typically between 0.2 and 0.5~m/s for differences between 1 and 4 pixels for the enlargment, which gives an estimate on the order of magnitude of the uncertainty. This estimation is conservative because part of this rms is caused by the interpolation, which is already taken into account. As the mask is increased, we also expect the uncertainty due to the interpolation to be larger than the 0.4 m/s uncertainty previously estimated. 

Finally, we note that the radial velocity measurements by MDI in regions where the magnetic field is strong may be  biased because of the use of a (quiet Sun) reference line to determine the velocity from the intensity measurements in the line, as lines are broadened by the Zeeman effect in magnetic regions. \cite{wachter06} found that the velocity may be underestimated by about 2\% for plages, and by up to 5\% in spot umbra. The effect on our reconstructed radial velocity is therefore expected to be small.

\subsection{Convective blueshift in magnetic structures}

The convective blueshift associated to the Ni line observed by MDI has never been studied. Therefore, these observations also constitute an opportunity of characterizing the velocity as a function of the local magnetic field in that particular case. This way, we also check the assumption made in Paper II for $\Delta$V, which was that granulation is abnormal in the network \cite[e.g. ][]{morinaga08}. 
For this specific study we considered image pairs separated by less than 12 minutes (so that the rotation is less than a pixel at disk center), to avoid the additional uncertainty that would be introduced by the rotation correction. This concerns 1901 image pairs. We considered two approaches, a study of the velocity in each pixel of magnetic regions, and a study of averaged velocities over all magnetic structures. In addition, only structures with $\mu$ (cosinus of the angle between the line-of-sight and 
the vertical to the surface) larger than 0.8 were considered here (except precised otherwise), i.e. very close to disk center, to eliminate as many projection effects as possible. 

We first defined magnetic regions using the threshold of Paper II, to be able to make a direct comparison (i.e. 100~G). We considered structures above 7 part-per-million of the solar disk (hereafter ppm). This led to about 18 millions pixels (10 millions for $\mu$ above 0.9). 
The 100~G threshold also provided 623808 structures (and 337808 for $\mu$ above 0.9).
We used a lower threshold (40~G) to study the behavior of weaker flux areas, leading to about 37 million pixels (21 millions for $\mu$ above 0.9), while considering structures above 4 ppm. 
In addition to the magnetic field, velocity field, and position of the pixel, we also retrieved the size of the region (defined according to the same threshold) that the pixel belongs to.

\section{Velocity in magnetic regions}

\subsection{Velocity versus local magnetic field}

\begin{figure} 
\includegraphics{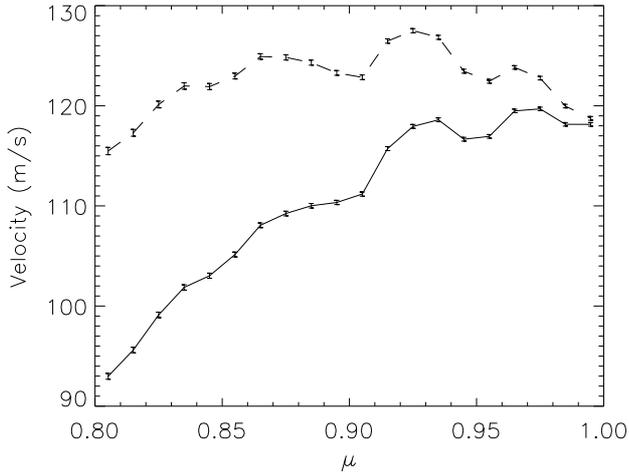}
\caption{Velocity (solid line) and velocity divided by $\mu$ (dashed line) as a function of $\mu$, in m/s, for magnetic fields larger than 100~G.}
\label{vmu}
\end{figure}

\begin{figure} 
\includegraphics{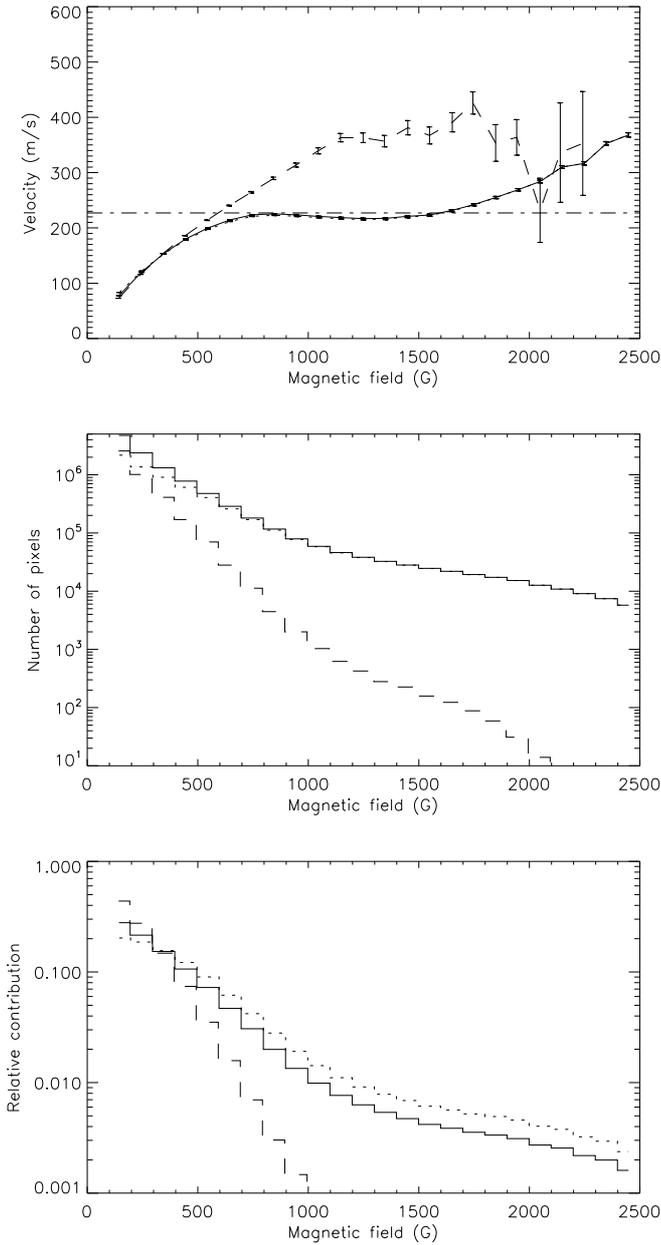}
\caption{{\it Upper panel}: $V/\mu$ versus the magnetic field, for $\mu>0.9$: for all pixels (solid line), pixels belonging to structures larger than 100~ppm (dotted line), and pixels belonging to structures smaller than 100~ppm (dashed line). The horizontal line correspond to 227 m/s (Paper II). 
{\it Middle panel}: Number of pixels versus the magnetic field for the same pixel selection. {\it Lower panel}: Relative contribution to the total RV. }
\label{vb}
\end{figure}

\subsubsection{Results}

Figure~\ref{vmu} shows the velocity $V$ and $V/\mu$ versus $\mu$, averaged over bins of 0.01 in $\mu$. After division by $\mu$ the velocity is expected to be constant if the velocity field is mostly vertical. The observed curve is quite flat, showing that we are mostly sensitive to vertical velocity fields. Strictly speaking, a portion of these pixels should also be associated to horizontal velocity fields, such as the Evershed effect (in active regions) or supergranular velocity field (in the network). Their contribution should be small, however, because we selected pixels very close to the disk center. Below we consider $V/\mu$ only, refered to as $V$ hereafter for simplicity.  
The velocity is positive, indicative of a redshift, which is consistent with an attenuation of the convective blueshift. 
 
Figure~\ref{vb} (solid line) shows the dependence of the velocity on the magnetic field for $\mu>$0.9, averaged over all pixels with the magnetic field in a certain range. We observe a stronger velocity field 
for larger magnetic fields, indicative of a stronger attenuation of the convective blueshift as expected. The average over the entire disk (for the 100~G threshold) is 123~m/s, which is lower than the expected 227~m/s derived from the work of \cite{bs90} and \cite{gray09} in Paper II. This value is used in the next section to compare our observations with the simulation.  The maximum value for the largest magnetic fields is close to the 340~m/s (convective blueshift for the Ni line derived from the assumptions made in Paper II) which would correspond to the total suppression of the convective blueshift for that line, which is indeed expected for umbral regions. The percentage of pixels with these strong fields is lower than the proportion of pixels expected in umbral area (a few percent of all pixels, excluding the network), so we expect a significant proportion of the pixels with a strong RV to belong to umbral regions. The amplitude of the curve presents a factor $\sim$4 only between the small and large magnetic fields (while these vary by a factor 25). 

When considering pixels down to 40~G, the average velocity over the entire disk is 83~m/s.
Note that an average velocity of 227 m/s corresponds to pixels with a magnetic field larger than 500 G. 
The lower panel of Fig.~\ref{vb}  shows that weak fields contribute very significantly to the total RV. 

\subsubsection{Convective blueshift attenuation versus downflows}

It is well known that the network magnetic structures can be associated to downflows, both at the granular scale (in intergranules) and at the supergranular scale (in regions of strong converging flows, outlining supergranular cells). Part of what we measure in the network could therefore be due to an actual downflow (which should be taken into account when computing the zero) and not to the attenuation of the convective blueshift. This downflow would also produce a redshift. Because of the poor resolution of full-disk images obtained by MDI (2 arcseconds, i.e. similar to a typical granule), the small-scale contribution should be negligible because we average the velocity over granules. 
We now investigate the possibility that part of the measured velocity associated to magnetic regions could be caused by an actual downflow in supergranules.

It is difficult to observationally make the difference between an actual downflow and an attenuation of the convective blueshift. 
\cite{miller84} for example observed velocities associated to network structure. They found that the three Fe lines they studied were redshifted in the network by 75--200~m/s (in excellent agreement with our results), but only when measured from the line wings. The signal was not present in the line core, which they interpreted as a sign that this redshift was not caused by an actual downflow, but by an attenuation of the convective blueshift (which leads to line distorsions). 
Several results have been obtained by various groups \cite[see the review by][]{solanki93}, and they are generally consistent with a modification of granulation due to magnetic fields. 

To investigate this question we here selected pixels according to the size of the region they belong to, because we do not expect any significant downflow bias in active regions due to the altered supergranulation there. 
This is shown in Fig.~\ref{vb}. We expect pixels in the network (i.e. small structures) to have a higher velocity 
(because they are predominantly in a downward flow which is superimposed on the redshift produced by the attenuation of the convective blueshift) compared to pixels with the same magnetic field in larger structures. We find that this is true only for pixels with a relatively large magnetic field of at least a few 100~G. These pixels represent only a few \% of all network pixels, so we consider that this effect can be neglected in our analysis. 
Also, the velocity - magnetic field relationship starts to change for structures below $\sim$200~ppm: these are already large structures, for which it would be difficult to have all pixels associated to an actual downflow.  
We also point out that due to the large mask, the local downflows should not significantly impact the estimation of the zero velocity at each time.

We conclude that most of the velocity amplitude we measure is probably due to the attenuation of the convective blueshift, in agreement with \cite{miller84}.

\subsection{Structure analysis}

\begin{figure} 
\includegraphics{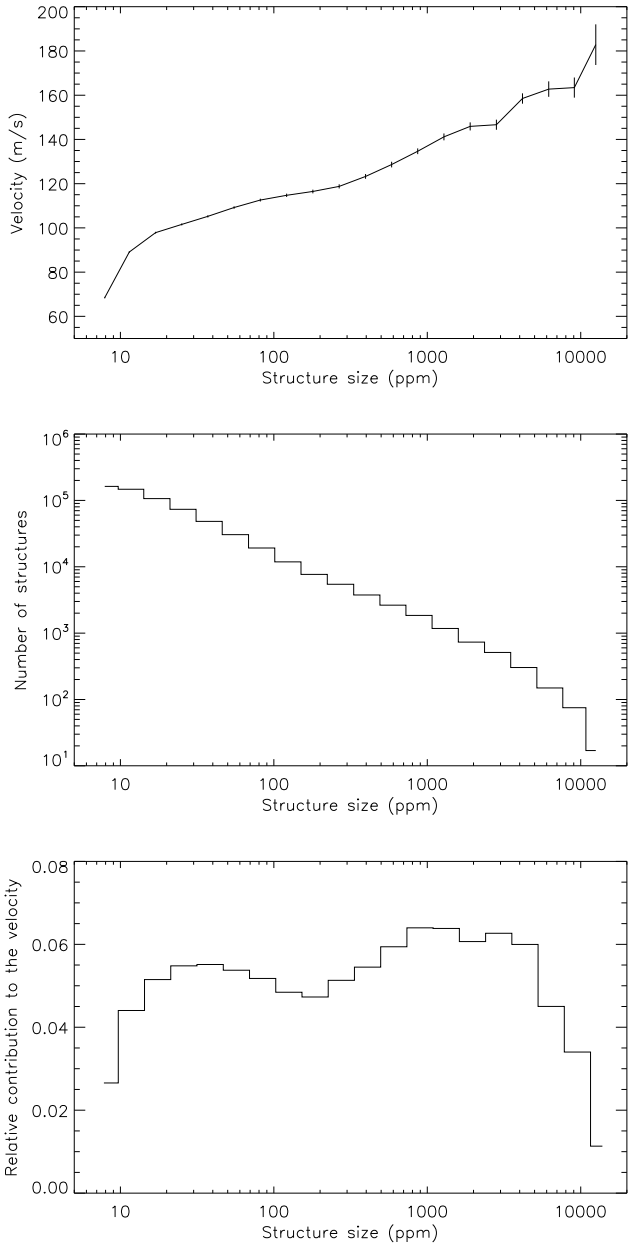}
\caption{{\it Upper panel}: Velocity averaged over the structure versus the structure size for the 100~G threshold structures. {\it Middle panel}: Number of structures versus the size. {\it Lower panel}: Relative contribution of each category of structures to the total RV versus the size.}
\label{struc}
\end{figure}

Figure~\ref{struc} (upper panel) shows the velocity field averaged over each magnetic structure as defined in Paper II. The velocity scales with the size of the structures, as expected from the larger contribution of strong magnetic field pixels. Note the change in slope at 300 ppm. This dependence of the RV on the size has not been taken into account in our simulation so far. This will be done in a forthcoming paper.

The bottom panel of Fig.~\ref{struc} also shows the relative contribution to the total velocity field for each category of structures. This relative contribution is the product defined by $N \times S \times V$ (normalized so that the sum of all contributions is equal to 1), where $N$ is the number of structures of size $S$ (middle panel) and $V$ the velocity averaged over the structures of size $S$ (upper panel). The largest contribution comes from large structures (67\% for structures larger than 100~ppm of the solar disk, i.e. $\sim$150~Mm$^2$). Small structures still contribute significantly however: for example, structures smaller than 26~ppm \cite[about 80~Mm$^2$, typical of the quiet network,][]{wang88} represent 14\% of the total velocity.  
The shape of the relative contribution into two components (for sizes above and below 200~ppm respectively) is related to the size distribution (middle panel of Fig.~\ref{struc}). Although the effect is small, there is a change in slope around 200~ppm of this size distribution, which creates the gap for this size: the product $S \times N$ also shows that shape. 

\section{Integrated radial velocity: results and analysis}

\begin{figure} 
\includegraphics{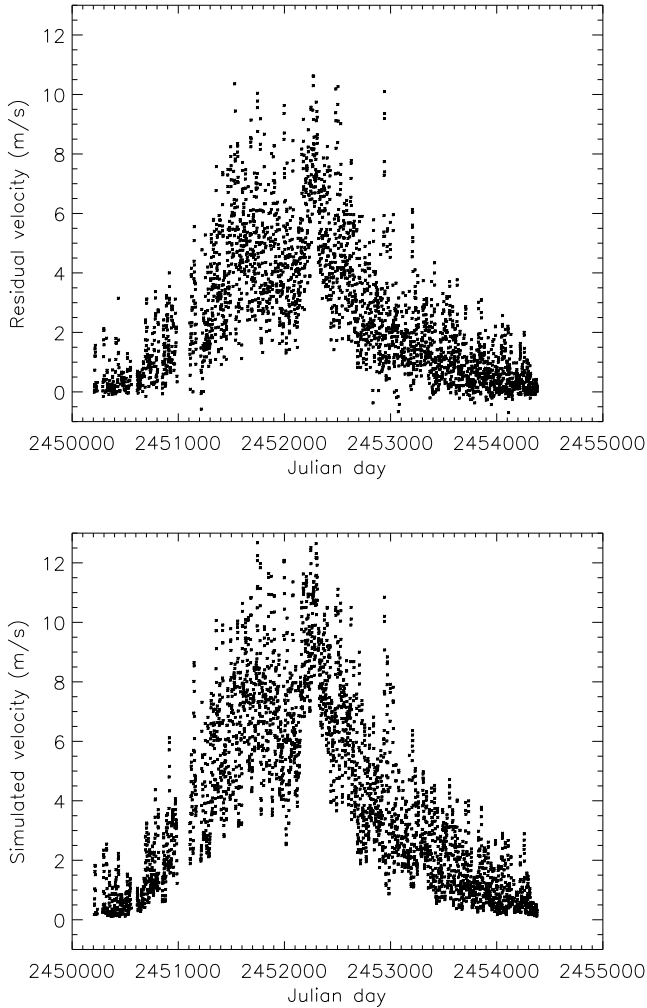}
\caption{ {\it Upper panel}: Residual velocity ($V_{\rm obs}$-$V_{\rm zero}$). {\it Lower panel}: Simulated RV variations due to the convective blueshift attenuation (from Paper II)  scaled to a $\Delta$V of 227~m/s.}
\label{vall2}
\end{figure}

\begin{figure} 
\includegraphics{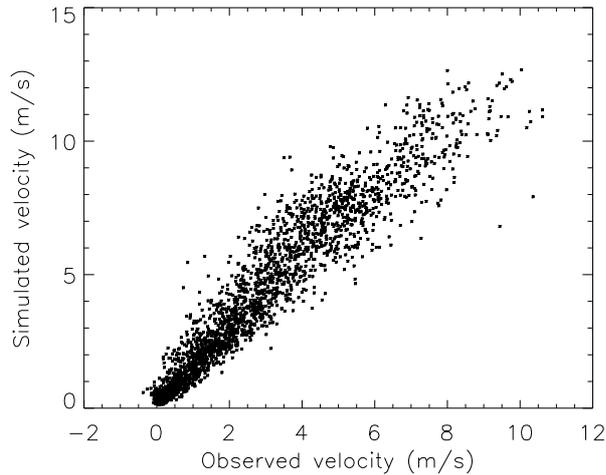}
\caption{Simulated velocity (from Paper II, scaled to 227~m/s) versus the observed velocity, whenever Dopplergrams and simulations are separated in time by less than one hour. }
\label{vcorr}
\end{figure}

\begin{figure} 
\includegraphics{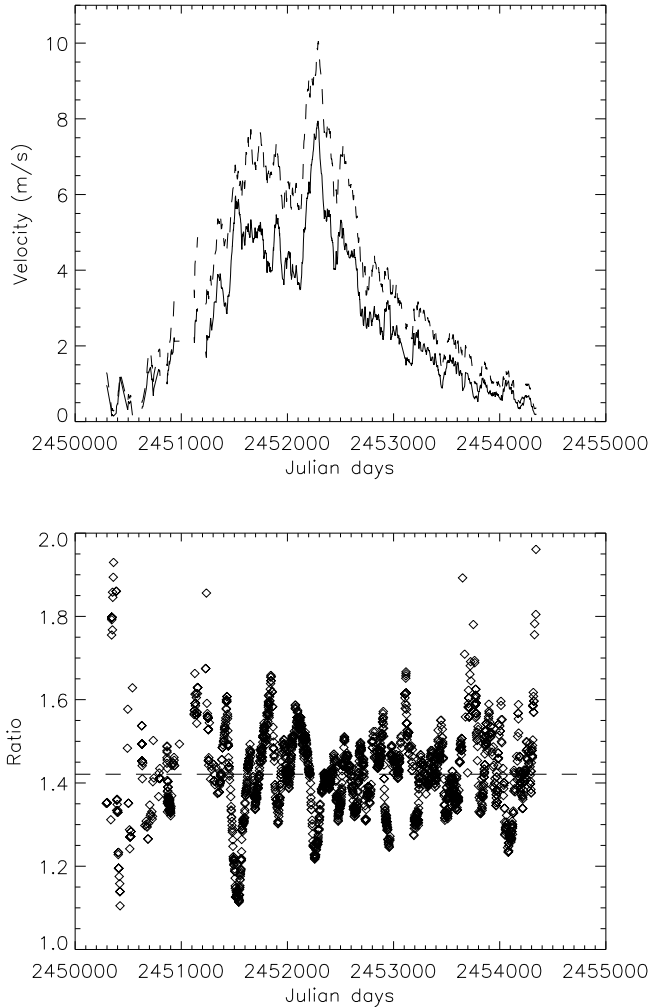}
\caption{{\it Upper panel}: Simulated RV smoothed over 30 days (from Paper II) scaled to 227~m/s (dashed line) and observation (solid), whenever Dopplergrams and simulations are separated in time by less than one hour. {\it Lower panel}: Ratio between the two (simulation divided by observation). The horizontal dashed line represents the median value (1.42).}
\label{vcomp1}
\end{figure}

\begin{figure} 
\includegraphics{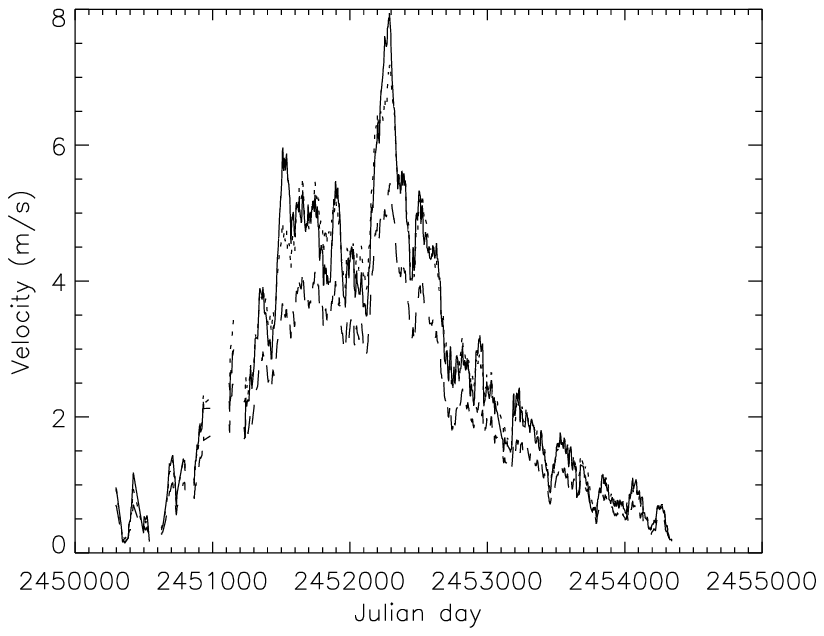}
\caption{Smoothed simulation (from Paper II) scaled to 123 m/s (dashed line) and to 166 m/s (dotted line), and observation (solid line), whenever Dopplergrams and simulations are separated in time by less than one hour. }
\label{vcomp2}
\end{figure}

\subsection{General results}

Figure~\ref{vall2} shows the complete time series of the deduced RV signal $V_{\rm obs}-V_{\rm zero}$ together with the simulated RV variations scaled to 227~m/s (we recall that this value is derived from the assumptions made in Paper II). 
The periodogram of the whole time series 
is very similar to the periodogram for the simulation computed in Paper II (see the bottom of their Fig.~10) and therefore variations on various time-scales agree well. 
Figure~\ref{vcorr} shows these simulated RV as a function of the observed ones, whenever Dopplergrams and simulations are separated in time by less than one hour (corresponding to 2328 points). They are correlated well (correlation of 0.94). The observed 
long-term variations are significantly above the uncertainties we estimated in Sect. 2.

Note that the observations show very few values below zero (less than 4\%), out of which 80\% are above -0.2~m/s. This is compatible with our uncertainty estimate made in Sect.~2.3.

\subsection{Long-time scale variations}

We now investigate the long-term variations of the RV. Below, we smooth the time series over 30 days to facilitate the comparison of the various levels, and focus on the long-term variations (i.e. above the solar rotation period). The smoothing is performed using a running mean (on these plots a boxcar is used, but the use of triangular mean gives similar results). Figure~\ref{vcomp1} 
compares the observed RV with the results of the simulation obtained in Paper II for the convection and scaled to 227~m/s, once smoothed over 30 days. The two curves are quite similar within 30\%. The observed long-term amplitude 
is about 70\% of the amplitude obtained in the simulation, which is a good consensus. It shows that our simulation produces the correct order of 
magnitude. The variations at various scales are also well reproduced.

We now compare the values deduced from our simulation scaled to $\Delta$V derived from our present MDI analysis (Sect.~3.1). Figure~\ref{vcomp2} shows 
the observations compared to the simulation scaled to 123~m/s (this corresponds to the average velocity 
for structures defined as in Paper II) instead of 227 m/s as in Figure~\ref{vcomp1}. Our simulation then predicts a smaller radial velocity signal than what is observed. 
The difference is because we considered in Paper II only pixels above 100 G. However, pixels below 100~G also contribute to the signal. This is shown by the dotted line, representing the RV simulated in Paper II scaled to 83~m/s 
and to the correct filling factor for this threshold of 40~G (there is a factor 2 between the number of pixels above 100 G 
and the number of pixels above 40 G): it is very close to the observed velocity. 

In conclusion, the simulation 
in Paper II overestimated the $\Delta$V that should be taken into account for the considered structures by about 30\%; 
but it underestimated the surface covered by magnetic regions. 

Finally, the $\Delta$V of 227 m/s represented 2/3 of the expected convective blueshift for that line (340 m/s, derived from the Ni line depth). The simulation would 
reproduce the RV for the structures defined by the 100 G threshold for an average $\Delta$V of 70\% of the 227 m/s. 
This corresponds to an average attenuation of the convective blueshift for these structures by a factor  0.47 (instead of 0.67). 
We insist that the actual value of the $\Delta$V to be used on a given data set of magnetic structures depends on their definition, because there is a crosstalk between $\Delta$V and the size of the structures (only the product is important).

\subsection{Small-time scale variations}

In addition to this general agreement for the long-term variations, it is interesting to compare the behavior of the observed and simulated short-term variations. 
We consider now the residuals after substraction of the smoothed time series defined in the previous section : the rms is 1.14 m/s for the observation, compared to 1.26 m/s for the simulation. If we multiply this last rms by 0.7 to take into account the difference in scale between observations and the simulation as shown in the previous section, the rms associated to the simulation becomes 0.88~m/s. Taking into account the uncertainties due to the interpolation (0.4~m/s), the quadratic difference leads to an extra rms for the observations of 0.6~m/s. It is confirmed by the rms of the observation minus the simulation scaled by 70\%, which leads to an rms of 0.75~m/s (and the quadratic difference with 0.4~m/s also gives 0.6~m/s). The additionnal uncertainty due to the mask definition is less precise, but a conservative estimate shows that it lies between 0.2 and 0.5~m/s. It could therefore be compatible with the remaining 0.6~m/s, although a small additionnal dispersion may be present. This means that our simulations reproduce the observations on small-time scales quite well, but also that other effects not taken into account in our simulation such as supergranulation motions and the Evershed effect contribute very little. Their contribution is therefore much lower than that of the convective blueshift attenuation.

\section{Conclusion}

We reconstructed the integrated RVs of the Sun from MDI Dopplergrams using the Ni line at 6768 \AA$\:$ over a solar cycle. We obtained a long-term amplitude of the variations of about 8~m/s. The resulting RV variations agree very well with the simulated RV (Paper II) once scaled to the Ni line: the small-scale and long term variations are well reproduced within 30\%. We therefore validate the simulations made in Paper II.  

We also analyzed in detail the variations of the relative velocity (with respect to the quiet Sun) in magnetic regions, which we interpreted as an attenuation of the convective blueshift in most cases. We characterized this velocity as a function of the local magnetic field and as a function of the structure size. Our results are compatible with an average attenuation of the convective blueshift in plages and network regions by about 50 \%, compared to a factor 2/3 in \cite{bs90} for plages. The difference probably arises because we included weaker field regions in our analysis.  

The variations of the RV due to the convective blueshift attenuation versus the structure size could be implemented in the simulations like that made in Paper II, because the structure size is one of the input parameters defining magnetic regions. The dependence on the local magnetic field is important to understand and test our RV variations. However, if we want to extrapolate this work to other stars, this relation may not be easy to take into account, because the magnetic field distribution on other stars is not well characterized.

\begin{acknowledgements}
SOHO is a mission of international cooperation between the European Space
Agency (ESA) and NASA. 
\end{acknowledgements}

\bibliographystyle{aa}
\bibliography{biblio14199}

\end{document}